\documentclass{ws-procs9x6}

\newcommand{\lsim}{\rlap{\raise 2pt \hbox{$<$}}{\lower 2pt
\hbox{$\sim$}}\ }
\newcommand{\gsim}{\rlap{\raise 2pt \hbox{$>$}}{\lower 2pt
\hbox{$\sim$}}\ }
\newcommand{\be}{\begin{equation}}
\newcommand{\ee}{\end{equation}}
\newcommand{\dslash}{\rlap{/}{\partial}}
\newcommand{\Aslash}{\rlap{/}{A}}

\newcommand{\s}{{\rm s}}

\begin{document}

\title{NEUTRINO PHENOMENOLOGY}

\author{ESTEBAN ROULET\footnote{\uppercase{W}ork partially
supported by \uppercase{F}undaci\'on \uppercase{A}ntorchas.}}

\address{CONICET, Centro At\'omico Bariloche\\
Av. Bustillo 9500, 8400, Bariloche, Argentina\\ 
E-mail: roulet@cab.cnea.gov.ar}

\maketitle

\abstracts{
A general overview of neutrino physics  is given, 
starting with a historical account of the
development of our understanding of neutrinos and how they helped to
unravel the structure of the Standard Model. We discuss why it is so
important to establish if neutrinos are massive and   the
indications in favor of non-zero neutrino masses are discussed, including
the recent results on atmospheric and solar neutrinos and their confirmation
with artificial neutrino sources.}

\section{The neutrino story:}

\subsection{The hypothetical particle:}

One may trace back the appearance of neutrinos in physics to the 
discovery of radioactivity by Becquerel one century ago. When the
energy of the electrons (beta rays) emitted in a radioactive decay
 was measured by Chadwick in 1914, it turned out to his surprise 
to be continuously distributed. 
This was not to be expected if the underlying process
in the beta decay was the transmutation of an element $X$ into 
another one $X'$ with the emission of an electron, i.e. $X\to X'+e¯$, 
since in that case the electron should be monochromatic. The situation 
was so puzzling that Bohr even suggested that the conservation
of energy may not hold in the weak decays. Another serious problem
with the `nuclear models' of the time was the belief that
nuclei consisted of protons and electrons, the only known particles 
by then. To explain the mass and the charge of a 
 nucleus it was then necessary that it 
had $A$ protons and $A-Z$ electrons in it. For instance,
a $^4$He nucleus would have
 4 protons and 2 electrons. Notice that this total of 
six fermions would make the $^4$He nucleus to be a boson, which 
is correct. However, the problem arouse when this theory was applied
for instance to $^{14}$N, since consisting of 14 protons and 7 electrons
would make it a fermion, but the measured angular momentum of the
nitrogen nucleus was $I=1$.

The solution to these two puzzles was suggested by Pauli only in 1930, 
in a famous letter to the `Radioactive Ladies and Gentlemen' 
gathered in a meeting in Tubingen, where he wrote: `I have hit 
upon a desperate remedy to save the exchange theorem of statistics 
and the law of conservation of energy. Namely, the possibility that 
there could exist in nuclei electrically neutral particles, that I wish to 
call neutrons, which have spin 1/2 ...'. These had to be not heavier
than electrons and interacting not more strongly than gamma rays.

With this new paradigm, the nitrogen nucleus became 
$^{14}$N$=14 p+7e+7`n$', which is a boson, 
and a beta decay now involved the emission of two particles
$X\to X'+e+`n$', and hence the electron spectrum was continuous. 
Notice that no particles were created in a weak
decay, both the electron and Pauli's neutron $`n$' were already 
present in the nucleus of the element $X$, and they just came 
out in the decay.
However, in 1932 Chadwick discovered the real `neutron', with a mass 
similar to that of the proton and being the missing building block
of the nuclei, so that a nitrogen nucleus finally became just
$^{14}$N$=7p+7n$, which also had  the correct bosonic statistics.

In order to account now for the beta spectrum of weak decays, 
Fermi called Pauli's hypotetised particle the neutrino 
(small neutron), $\nu$, 
and furthermore suggested that the fundamental process underlying
beta decay was $n\to p+e+\nu$. He wrote \cite{fe34} 
the basic interaction by analogy with the interaction known at the time,
the QED, i.e. as a vector$\times$vector current interaction:

$$H_F=G_F\int {\rm d}^3x[\bar\Psi_p\gamma_\mu\Psi_n]
[\bar\Psi_e\gamma^\mu\Psi_\nu]+ h.c. .$$
This interaction accounted for the continuous beta spectrum, and 
 from the measured shape at the endpoint Fermi concluded that
$m_\nu$ was consistent with zero and had to be small.
The Fermi coupling $G_F$ was estimated from the observed lifetimes 
of radioactive elements, and armed with this Hamiltonian 
Bethe and Peierls \cite{be34} decided to compute the cross section
for the inverse beta process, i.e. for $\bar\nu+p\to n+e^+$, which was
the relevant reaction to attempt the direct detection of a neutrino. 
The result, $\sigma=4(G_F^2/\pi)p_eE_e\simeq 2.3\times 10^{-44}$cm$^2
(p_eE_e/m_e^2)$ was so tiny that they concluded `... This meant that
one obviously would never be able to see a neutrino.'. For instance, 
if one computes the mean free path in water (with density 
$n\simeq 10^{23}/$cm$^3$) 
of a neutrino with energy $E_\nu=2.5$ MeV, typical of a weak decay, 
the result is $\lambda\equiv 1/n\sigma\simeq 2.5\times 10^{20}$ cm, 
which is $10^7$AU, i.e. comparable to the thickness of the 
Galactic disk.

It was only in 1956 that Reines and Cowan  were able to prove
that Bethe and Peierls had been too pessimistic, when they 
measured for the first time the interaction of a neutrino 
through the inverse beta process\cite{re59}. Their strategy was 
essentially that, if one needs $10^{20}$ cm of water to stop a neutrino, 
 having $10^{20}$ neutrinos a cm would be enough to stop one neutrino.
Since after the second war powerful reactors started to become available,
and taking into account that in every fission of an uranium
nucleus the neutron rich fragments beta decay producing typically
6 $\bar\nu$ and liberating $\sim 200$ MeV, it is easy to show that
the (isotropic) neutrino flux at a reactor is
$${d\Phi_{\nu}\over d\Omega}\simeq {2\times 10^{20}\over 4\pi}
\left( {{\rm Power\over GWatt}}\right){\bar\nu\over strad}.$$  
Hence, placing a few hundred liters of water (with some Cadmium in it)
 near a reactor they
were able to see the production of positrons (through the two 
511 keV $\gamma$ produced in their annihilation with electrons) and 
neutrons (through the delayed $\gamma$ from the neutron capture in Cd), 
with a rate consistent with the expectations from the weak interactions
of the neutrinos.

\subsection{The vampire:}

Going back in time again to follow the evolution of the 
theory of weak interactions of neutrinos, in 1936 Gamow and Teller
\cite{ga36} noticed that the $V\times V$ Hamiltonian of Fermi was
probably too restrictive, and they suggested the generalization

$$H_{GT}=\sum_iG_i[\bar pO_in][\bar eO^i\nu]+h.c. ,$$
involving the operators $O_i=1,\ \gamma_\mu,\ \gamma_\mu\gamma_5,\ 
\gamma_5,\ \sigma_{\mu\nu}$, corresponding to scalar ($S$), 
vector ($V$), axial vector $(A)$, pseudoscalar ($P$) and tensor ($T$)
currents. However, since $A$ and $P$ only appeared here as $A\times A$ or 
$P\times P$, the interaction was parity conserving.
The situation became unpleasant, since now there were five different
coupling constants $G_i$ to fit with experiments, but however
this step was required since some observed 
nuclear transitions which were
forbidden for the Fermi interaction became now allowed with its
generalization (GT transitions). 

The story became  more involved when in 1956 Lee and Yang
suggested that parity could be violated in weak
interactions\cite{le56}.  This 
could explain why the particles theta and tau had exactly the
same mass and charge and only differed in that the first one 
was decaying to two pions while the second to three pions (e.g. to
states with different parity). The explanation to the 
puzzle was that the $\Theta$ and $\tau$  were 
just the same particle, now known as the charged kaon, but the 
(weak) interaction leading to its decays violated parity.

Parity violation was confirmed the same year by Wu \cite{wu57}, 
studying the direction of emission of the electrons emitted in
the beta decay of polarised $^{60}$Co. The decay rate is
 proportional to $1+\alpha \vec{P}\cdot \hat{p}_e$. 
Since the Co polarization vector $\vec P$ is an axial vector, while the
unit vector along the electron momentum $\hat{p}_e$ is a vector, their
scalar product is a pseudoscalar and hence a non--vanishing 
coefficient $\alpha$ would imply parity violation. The result was 
that electrons preferred to be emitted in the direction opposite 
to $\vec{P}$, and the measured 
value $\alpha\simeq -0.7$ had then profound implications for
the physics of weak interactions. 

The generalization by  Lee and Yang of the Gamow Teller Hamiltonian
was
$$H_{LY}=\sum_i[\bar pO_in][\bar eO^i(G_i+G_i'\gamma_5)\nu]+h.c. .$$
Now the presence of terms such as $V\times A$ or $P\times S$ allows
for parity violation, but clearly the situation became even more 
unpleasant since there are now 
10 couplings ($G_i$ and $G_i'$) to determine, so that some order was
really called for.

Soon the bright people in the field realized that there could be
a simple explanation of why parity was violated in weak interactions,
the only one involving neutrinos, and this had just to do with
the nature of the neutrinos. Lee and Yang, Landau and Salam
\cite{bright} realized that, if the neutrino was
massless, there was no need to have both neutrino chirality states 
in the theory, and hence the handedness of the neutrino could be the
origin for the parity violation. 
 To see this, consider the chiral projections of a fermion
$$\Psi_{L,R}\equiv {1\mp \gamma_5\over 2}\Psi.$$
We note that   
in the relativistic limit these two projections describe
left and right handed helicity states (where the helicity, i.e. the
spin projection in the direction of motion, is a constant of motion
for a free particle), but in general an helicity eigenstate is a
mixture of the two chiralities. For a massive particle, which
has to move 
 with a velocity smaller than the speed of light, it is always
possible to make a boost to a system where the helicity is reversed, 
and hence the helicity is clearly not a Lorentz invariant while the
chirality is (and hence has the desireable properties of a charge to
which a gauge boson can be coupled).
If we look now to the equation of motion for a Dirac particle
 as the one we are used to for the description of
a charged massive particle such as an electron ($(i\dslash-m)\Psi=0$), 
 in terms of the chiral projections this equation becomes
$$i\dslash\Psi_L=m\Psi_R$$
$$i\dslash\Psi_R=m\Psi_L$$
and hence clearly a mass term will mix the two chiralities.
However, from these equations we see that for $m=0$, as could be the
case for the neutrinos, the two equations are decoupled, and one 
could write a consistent theory using only one of 
the two chiralities (which in this case would coincide with the
helicity).  If the Lee Yang Hamiltonian were just to depend
on a single neutrino chirality, one would have then $G_i=\pm G_i'$ and 
parity violation would indeed be
maximal. This situation has been described by
saying that neutrinos are like vampires in Dracula's stories: 
if they were to look to
themselves into a mirror they would 
be unable to see their reflected images.

The actual helicity of the neutrino was measured
by Goldhaber et al. \cite{go58}.
The experiment consisted in observing the $K$-electron capture
in $^{152}$Eu ($J=0$) which produced  $^{152}$Sm$^*$ ($J=1$) 
 plus a neutrino. This excited nucleus then decayed into 
$^{152}$Sm ($J=0)+\gamma$. Hence the measurement of the polarization 
of the photon gave the required 
information on the helicity of the neutrino 
emitted initially. The conclusion was that `...Our results
seem compatible with ... 100\% negative helicity for the neutrinos', 
i.e. that the neutrinos are left handed particles.

This paved the road for the $V-A$ theory of weak interactions
advanced by Feynman and Gell Mann, and Marshak and 
Soudarshan \cite{vma}, which stated that weak interactions only
involved vector and axial vector currents, in the combination 
$V-A$ which only allows the coupling to left handed fields, i.e.
$$J_\mu=\bar e_L\gamma_\mu\nu_L+\bar n_L\gamma_\mu p_L$$
with $H=(G_F/\sqrt{2})J_\mu^\dagger J^\mu$.
This interaction also predicted the existence of purely leptonic weak 
charged currents, e.g. $\nu+e\to \nu+e$, to be experimentally 
observed much later\footnote{A curious fact was that the new theory
predicted a cross section for the inverse beta decay a factor of two
larger than the Bethe and Peierls original result, which 
had already been  confirmed in 1956 to the 5\% accuracy by 
Reines and Cowan. However, in an improved 
 experiment in 1959 Reines and Cowan found 
 a value consistent with the new
prediction, what shows that many times when the experiment agrees with
the theory accepted at the moment the errors tend to be underestimated.}. 

The current
involving  nucleons is actually 
not exactly $\propto \gamma_\mu(1-\gamma_5)$
(only the interaction at the quark level has this form), but
is instead $\propto \gamma_\mu(g_V-g_A\gamma_5)$. The vector
coupling remains however
 unrenormalised ($g_V=1$) due to the so called conserved
vector current hypothesis (CVC), which states that the vector part of the
weak hadronic charged currents ($J_\mu^\pm\propto 
\bar \Psi\gamma_\mu \tau^\pm\Psi$, with $\tau^\pm$ the raising 
and lowering operators in the isospin space $\Psi^T=(p,n)$) together with
the isovector part of the electromagnetic current (i.e. the term 
proportional to $\tau_3$ in the decomposition $J^{em}_\mu\propto
\bar \Psi\gamma_\mu (1+\tau_3)\Psi$) form an isospin triplet
of conserved currents. On the other hand, the axial vector hadronic
current is not protected from strong interaction renormalization effects
and hence $g_A$ does not remain equal to unity. The measured value,
using for instance the lifetime of the neutron, is $g_A=1.27$, so that
at the nucleonic level the charged current 
weak interactions are actually ``$V-1.27A$''.

With the present understanding of weak interactions, we know that
the clever idea  to explain parity 
violation as due to the non-existence of one of the neutrino
chiralities (the right handed one) was completely wrong, although
it lead to major advances in the theory and ultimately 
to the correct interaction. Today we understand that the parity violation 
is a property of the gauge boson (the $W$) responsible for the gauge 
interaction, which  couples only to the left handed fields, 
and not due to the
absence of right handed fields. For instance, in the quark sector both
left and right chiralities exists, but parity is violated because
the right handed fields are singlets for the weak charged currents.

\subsection{The trilogy:}

In 1947 the muon was discovered in cosmic rays by Anderson and Neddermeyer.
This particle was just a heavier copy of the electron, and as was
suggested by Pontecorvo, it also had weak interactions $\mu +p\to n+\nu$ 
with the same universal strength $G_F$. Hincks, Pontecorvo and Steinberger
showed that the muon was decaying to three particles, $\mu\to e\nu\nu$, 
and the question arose whether the two emitted neutrinos were 
similar or not. It was then shown by Feinberg \cite{fe58} that, assuming
the two particles were of the same kind, weak interactions couldn't be
mediated by gauge bosons (an hypothesis suggested in 1938 by Klein).
The reasoning was that if the two neutrinos were equal, it would be
possible to join the two neutrino lines and attach a photon to the
virtual charged gauge boson ($W$) or to the external legs\footnote{this
  reasoning would have actually also excluded a purely leptonic generalisation
  of a \uppercase{F}ermi's theory to describe the muon decay.}, 
so as to generate a diagram for
the radiative decay $\mu\to e\gamma$. The resulting branching ratio
would be larger than $10^{-5}$ and was 
hence already excluded at that time.
This was probably the first use of `rare decays' to constrain 
the properties of new particles.

The correct explanation for the absence of the radiative decay
was put forward by Lee and Yang, 
who suggested that the two neutrinos
emitted in the muon decay had different flavour, 
i.e. $\mu\to e+\nu_e+\nu_\mu$, and hence it was not possible 
to join the two neutrino lines to draw the  radiative decay diagram. 
This was confirmed at Brookhaven 
in the first accelerator neutrino experiment\cite{da62}. 
They used an almost
pure $\bar\nu_\mu$ beam,
something which can be obtained from charged 
pion decays, since the $V-A$ theory
implies that $\Gamma(\pi\to \ell+\bar\nu_\ell)\propto m_\ell^2$, i.e. 
this process requires a chirality flip in the final lepton line which
strongly suppresses the decays $\pi\to e+\bar\nu_e$.
Putting a detector in front of this beam they were able to observe
the process $\bar\nu+p\to n+\mu^+$, but no production of positrons, 
what proved that the neutrinos produced in a weak decay in association
with a muon were not the same as those produced in a beta
decay (in association with an electron).
Notice that although the neutrino fluxes are much smaller at 
accelerators than at reactors, their higher energies make their detection
feasible due to the larger cross sections ($\sigma\propto E^2$ 
for $E\ll m_p$, and $\sigma\propto E$ for $E\gsim m_p$).

In 1975 the $\tau$hird charged lepton was discovered by Perl at
SLAC, and being just
a heavier copy of the electron and the muon, it was concluded that
a third neutrino flavour had also to exist. The direct 
detection of the $\tau$ neutrino has been achieved by the
DONUT experiment at Fermilab, looking at the short $\tau$ tracks
produced by the interaction of a $\nu_\tau$ emitted in the decay of
 a heavy meson (containing a $b$ quark) produced in a beam dump.
Furthermore, we know today that
the number of light weakly interacting neutrinos is precisely three
(see below), so that the proliferation of neutrino species seems
to be now under control.

\subsection{The gauge theory:}

As was just mentioned, Klein had suggested that the short range
charged current weak interaction could be due to the exchange of
a heavy charged vector boson, the $W^\pm$. 
This boson exchange would look at
small momentum transfers ($Q^2\ll M_W^2$) as the non renormalisable 
four fermion interactions discussed before. If the gauge interaction
is described by the Lagrangian $\mathcal{L}=-(g/\sqrt{2})J_\mu W^\mu+h.c.$,
from the low energy limit one can identify the Fermi coupling as
$G_F=\sqrt{2}g^2/(8M_W^2)$.
In the sixties, Glashow, Salam and Weinberg showed that it was
possible to write down  a unified description of electromagnetic
and weak interactions with a gauge theory based in the group $SU(2)_L\times
U(1)_Y$ (weak isospin $\times$ hypercharge, with the electric charge
being $Q=T_3+Y$), 
which was spontaneously broken at the weak scale down 
to the electromagnetic $U(1)_{em}$. This (nowadays standard) model 
involves the three gauge bosons in the adjoint of $SU(2)$, $V_i^\mu$ (with
$i=1,2,3$), and the hypercharge gauge field $B^\mu$, so that the starting 
Lagrangian is 
$$\mathcal{L}=-g\sum_{i=1}^3J^i_\mu V^\mu_i-g'J^Y_\mu B^\mu + h.c. ,$$
with  $J^i_\mu\equiv \sum_a \bar \Psi_{aL}\gamma_\mu (\tau_i/2)\Psi_{aL}$.
The left handed leptonic and quark
isospin doublets are $\Psi^T=({\nu_e}_L,e_L)$ and 
($u_L, d_L)$ for the first generation (and similar ones for the other two
heavier generations) and the right handed fields are SU(2) singlets. 
The hypercharge current is obtained by summing 
over both left and right handed fermion chiralities and is
$J^Y_\mu\equiv \sum_a Y_a\bar \Psi_{a}\gamma_\mu \Psi_{a}$.

After the electroweak breaking one can identify the
weak charged currents with $J^\pm=J^1\pm iJ^2$, which couple to the
$W$ boson $W^\pm=(V^1\mp iV^2)/\sqrt{2}$, and the two neutral vector
bosons $V^3$ and $B$ will now combine through a rotation by the 
weak mixing angle $\theta_W$ (with tg$\theta_W=g'/g$), to give
\begin{eqnarray}
\begin{pmatrix}
 A_\mu \cr Z_\mu\end{pmatrix}=\begin{pmatrix} {\rm c}\theta_W & \s\theta_W\cr
-\s\theta_W & {\rm c}\theta_W \end{pmatrix} \begin{pmatrix} 
B_\mu\cr V^3_\mu
\end{pmatrix}.
\end{eqnarray}
We see that the broken theory has now, besides the massless photon field
$A_\mu$,  an additional neutral vector boson, the heavy $Z_\mu$, 
whose mass 
turns out to be related to the $W$ boson mass through 
$s^2\theta_W=1-(M_W^2/M_Z^2)$. The electromagnetic and neutral weak
currents are given by 
$$J_\mu^{em}=J^Y_\mu+J^3_\mu$$
$$J^0_\mu=J^3_\mu-\s^2\theta_W J^{em}_\mu,$$
with the electromagnetic coupling being $e=g\ \s\theta_W$.

The great success of this model came in 1973 with the experimental
observation of the weak neutral currents using muon neutrino beams
at CERN (Gargamelle) and Fermilab, using the elastic 
process $\nu_\mu e\to \nu_\mu e$. The semileptonic processes
$\nu N\to \nu X$ were also studied and the comparison of neutral
and charged current rates provided a measure of the weak mixing angle.
From the theoretical side t'Hooft proved the renormalisability of the
theory, so that the computation of radiative corrections became 
also meaningful.

\begin{figure}[t]
\centerline{\hbox{\epsfxsize=7cm \epsfbox{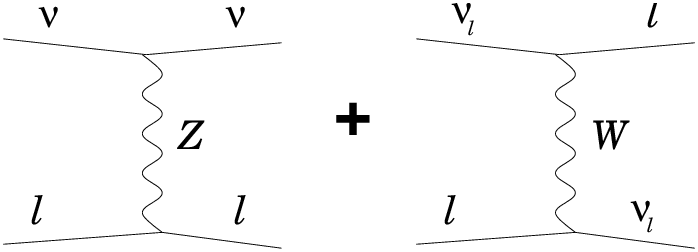} }}
\caption{\footnotesize 
Neutral and charged current contributions to neutrino 
lepton scattering.
\label{eps1}}
\end{figure}

The Hamiltonian for the leptonic weak interactions 
$\nu_\ell+\ell'\to\nu_\ell+\ell'$ can be obtained, 
using the Standard Model just presented, from the two diagrams 
in figure 1. In the low energy limit ($Q^2\ll M_W^2,\ M_Z^2$), 
it is just given by
$$H_{\nu_\ell\ell'}=\sqrt{2} G_F[\bar\nu_\ell \gamma_\mu
(1-\gamma_5)\nu_\ell][\bar\ell'\gamma^\mu(c_LP_L+c_RP_R)\ell']$$
where the left and right couplings are $c_L=\delta_{\ell\ell'}+
s^2\theta_W-0.5$ and $c_R=s^2\theta_W$. The $\delta_{\ell\ell'}$ term 
in $c_L$ is due to the charged current diagram, which clearly only
appears when $\ell=\ell'$. On the other hand, one sees that
due to the $B$ component in the $Z$ boson, the weak neutral currents
also  couple to the charged lepton right handed chiralities 
(i.e. $c_R\neq 0$).
This interaction leads to the cross section (for $E_\nu\gg m_{\ell'}$)
$$\sigma(\nu+\ell\to \nu+\ell)={2G_F^2\over \pi}m_\ell 
E_\nu\left[c_L^2+{c_R^2 \over 3}\right],$$
and a similar expression with $c_L\leftrightarrow c_R$ for antineutrinos.
Hence, we have the following relations for the neutrino
 elastic scatterings off electrons 
$$\sigma(\nu_e e)\simeq 9\times 10^{-44}{\rm cm}^2\left(
{E_\nu\over 10\ {\rm MeV}}\right)\simeq 2.5\sigma(\bar\nu_e e)
\simeq 6\sigma(\nu_{\mu,\tau}e)\simeq 7\sigma(\bar\nu_{\mu,\tau}e).$$
Regarding the angular distribution of the electron momentum with 
respect to the incident neutrino direction, in the center
of mass system of the process $d\sigma(\nu_e e)/d\cos\theta\propto
1+0.1 [(1+\cos \theta)/2]^2$, and it is hence almost isotropic. However, 
due to the boost to the laboratory system, there will be a 
significant correlation between the neutrino and electron momenta
for $E_\nu\gg $MeV, and this actually allows 
to do astronomy with neutrinos.
For instance, water cherenkov detectors such as Superkamiokande detect
solar neutrinos using this process, and have been able to reconstruct
a picture of the Sun with neutrinos. It will turn also to be relevant
for the study of neutrino oscillations that these kind of detectors 
are six times more sensitive to electron type neutrinos than to the
other two neutrino flavours.

Considering now the neutrino nucleon interactions, one has at low
energies (1~MeV$<E_\nu<50$~MeV) the cross section for the quasi-elastic
 process\footnote{Actually for the $\bar{\nu}_e p$ CC  interaction, 
the threshold energy is $E_\nu^{th}\simeq m_n-m_p+m_e\simeq 1.8$~MeV.}
$$\sigma(\nu_en\to p e)\simeq \sigma(\bar\nu_ep\to n e^+)\simeq 
{G_F^2\over \pi}{\rm c}^2\theta_C(g_V^2+3g_A^2)E_\nu^2,$$
where we have now
introduced the Cabibbo mixing angle $\theta_C$ which relates,
 if we ignore the third family, 
the quark flavour eigenstates $q^0$ to the mass eigenstates $q$, 
i.e. $d^0={\rm c}\theta_C d+\s\theta_Cs$ and $s^0=-\s\theta_C
d+{\rm c}\theta_Cs$ (choosing a flavour basis so that  the 
up type quark flavour and mass eigenstates coincide).

\begin{figure}[t]
\centerline{\hbox{\epsfxsize=9cm \epsfbox{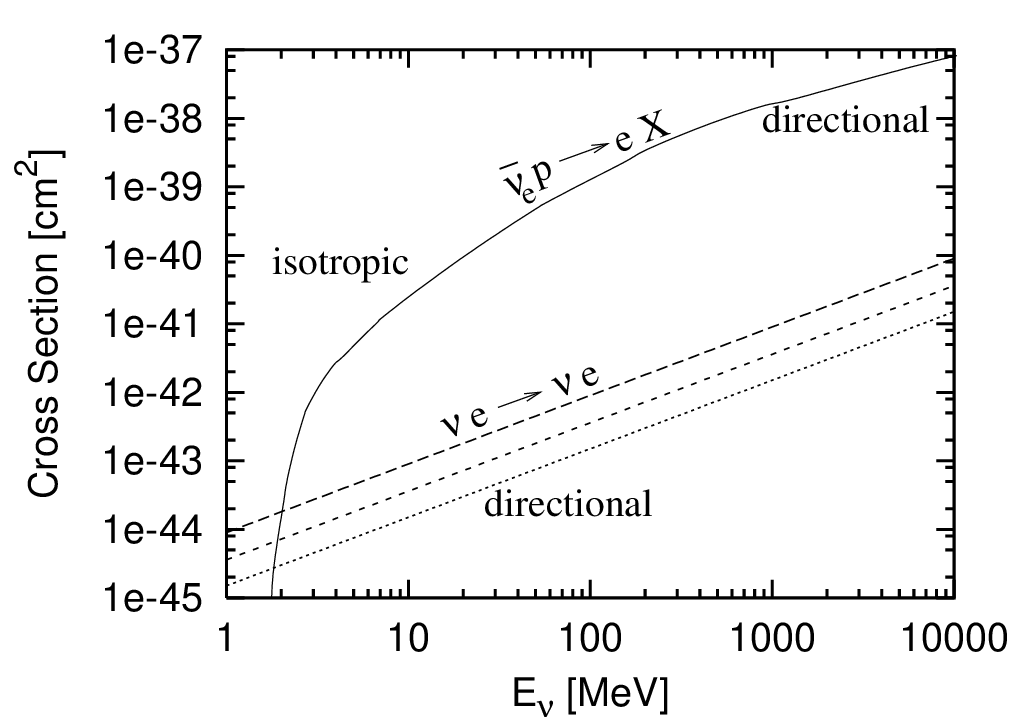} }}
\caption{\footnotesize 
Neutrino nucleon and neutrino lepton cross sections (the three 
lines correspond, from top to bottom, to the 
$\nu_e,\ \bar\nu_e$ and $\nu_{\mu,\tau}$ cross sections with electrons).
\label{sigma}}
\end{figure}

At $E_\nu\gsim 50$~MeV, 
the nucleon no longer looks like a point-like object
for the neutrinos, and hence the vector ($v_\mu$) and axial ($a_\mu$) 
hadronic currents
involve now momentum dependent form factors, i.e.
$$\langle N(p')|v_\mu|N(p)\rangle= \bar u(p')[
\gamma_\mu F_V+{i\over 2m_N}\sigma_{\mu\nu}q^\nu F_W] u(p)$$
$$\langle N(p')|a_\mu|N(p)\rangle= \bar u(p')[
\gamma_\mu \gamma_5F_A+{\gamma_5\over m_N}q_\mu F_P] u(p),$$
where $F_V(q^2)$ can be measured using electromagnetic processes 
and the CVC relation $F_V=F_V^{em,p}-F_V^{em,n}$ (i.e. as the difference
 between the proton and neutron electromagnetic 
vector form factors). Clearly $F_V(0)=1$ 
and $F_A(0)=1.27$, while $F_W$ is related to the anomalous magnetic moments
of the nucleons. The $q^2$ dependence has the effect of significantly 
flattening the cross section.
In the deep inelastic regime, $E_\nu\gsim$GeV, 
the neutrinos interact directly with the quark constituents. 
The cross section in this regime grows linearly with energy, and
this provided an important test of the parton model. 
The main characteristics of the neutrino cross section just discussed
are depicted in figure~2. For even larger energies, the gauge boson
propagators enter into the play (e.g. $1/M_W^2\to 1/q^2$) and the
growth of the cross section is less pronounced  above 10~TeV
($\sigma\propto E^{0.36}$). 

The most important test of the standard model came with the 
direct production
of the $W^\pm$ and $Z$ gauge bosons at CERN in 1984, and with the 
precision measurements achieved with the $Z$ factories LEP and SLC 
after 1989. These
$e^+e^-$ colliders working at and around the $Z$ resonance ($s=M_Z^2 =
($91~GeV$)^2$) turned out to be also crucial for neutrino physics,
since  studying the shape of the $e^+e^-\to f\bar f$ cross section
near the resonance, which has the Breit--Wigner form
$$\sigma\simeq{12\pi \Gamma_e\Gamma_f\over M_Z^2} {s\over (s-M_Z^2)^2+
M_Z^2\Gamma_Z^2},$$
it becomes possible to determine the total $Z$ width $\Gamma_Z$. This
width is just the sum of all possible partial widths, i.e.
$$\Gamma_Z=\sum_f\Gamma_{Z\to f\bar f}=\Gamma_{vis}+\Gamma_{inv}.$$
The visible (i.e. involving charged leptons and quarks) width
$\Gamma_{vis}$ can be measured directly, and hence one can infer a value 
for the invisible width $\Gamma_{inv}$. Since in the standard model
this last arises from the decays $Z\to \nu_i\bar\nu_i$, whose 
expected rate for decays into a given neutrino flavour is
$\Gamma_{Z\to \nu\bar\nu}^{th}=167$~MeV, one can finally 
obtain the number of neutrinos coupling to the $Z$ as $N_\nu=\Gamma_{inv}/
\Gamma_{Z\to \nu\bar\nu}^{th}$. The present best value for this quantity
is $N_\nu=2.994\pm 0.012$, giving then a 
strong support to the three generation standard model.

\bigskip

Going through the history of the neutrinos we have seen that
they have been extremely useful to understand the standard model.
On the contrary, the standard model is
of little help to understand the neutrinos. Since in the 
standard model there is no need
for $\nu_R$, neutrinos are  massless in this theory. There is however 
no deep principle behind this (unlike the masslessness of the photon
which is protected by the electromagnetic gauge symmetry), and indeed 
in many extensions of the standard model neutrinos turn out to be 
 massive. This 
makes the search for non-zero neutrino masses a very important
issue, since it provides a window to look for physics beyond the
standard model. Indeed, solid evidence has accumulated in the last years
indicating that neutrinos are massive, and this makes the field of neutrino
physics even more exciting now than in the long historic period that we have
just reviewed.

\section{Neutrino masses:}

\subsection{Dirac or Majorana?}

In the standard model, charged leptons (and also quarks) 
get their masses through their
Yukawa couplings to the Higgs doublet field $\phi^T=(\phi_+,\phi_0)$
$$-\mathcal{L}_Y=\lambda \bar L\phi\ell_R+h.c.\ ,$$
where $L^T=(\nu,\ell)_L$ is a lepton doublet and $\ell_R$ an SU(2)
singlet field. 
When the electroweak symmetry gets broken by the vacuum expectation 
value of the neutral component of the Higgs field $\langle \phi_0\rangle
=v/\sqrt{2}$ (with $v=246$~GeV), the following `Dirac' mass term results
$$-\mathcal{L}_m=m_\ell (\bar\ell_L\ell_R+\bar\ell_R\ell_L)=m_\ell\bar\ell
\ell,$$
where $m_\ell=\lambda v/\sqrt{2}$ and $\ell=\ell_L+\ell_R$ is the Dirac 
spinor field. This mass term is clearly invariant under the $U(1)$
transformation $\ell\to {\rm exp}(i\alpha)\ell$, which 
corresponds to the lepton number (and actually in this case also
to the electromagnetic gauge invariance). From 
the observed fermion masses, one concludes that the Yukawa couplings 
range from $\lambda_t\simeq 1$ for the top quark up to $\lambda_e
\simeq 10^{-5}$ for the electron.

Notice that the mass terms always couple fields with 
opposite chiralities, i.e. requires a $L\leftrightarrow R$ transition.
Since in the standard model the right handed neutrinos are not 
introduced, it is not possible to write a Dirac mass term, and hence
the neutrino results massless. Clearly the simplest way to give the
neutrino a mass would be to introduce the right handed fields just
for this purpose (having no gauge interactions, these sterile states 
would be essentially undetectable and unproduceable). 
Although this is a logical possibility, it has the 
ugly feature that in order to get reasonable neutrino masses, below the 
eV, would require unnaturally small Yukawa couplings ($\lambda_\nu 
<10^{-11}$). Fortunately it turns out that neutrinos are also very
special particles in that, being neutral, there are other ways 
to provide them a mass. Furthermore, in some scenarios
 it becomes also possible to get a natural
understanding of why neutrino masses are so much smaller than the charged
fermion masses.

The new idea is that the left handed neutrino field actually involves
two degrees of freedom, the left handed neutrino associated with the
positive beta decay (i.e. emitted in association with a positron) and
the other one being  the right handed
`anti'-neutrino emitted in the negative beta decays 
(i.e. emitted in association with an electron). 
It may then be possible to write down
a mass term using just these two degrees of freedom and involving the 
required $L\leftrightarrow R$ transition. This possibility was first 
suggested by Majorana in 1937, in a paper named `Symmetric theory of 
the electron and positron', and devoted mainly to the problem of 
getting rid of the negative energy sea of the Dirac
equation\cite{ma37}.  As
a side product, he found that for neutral particles there was `no more
any reason to presume the existence of antiparticles', and that `it was 
possible to modify the theory of beta emission, both positive and 
negative, so that it came always associated with the emission of a 
neutrino'. The spinor field associated to this formalism was then
named in his honor a Majorana spinor.

To see how this works it is necessary to introduce the so called 
antiparticle field, $\psi^c\equiv C\bar\psi^T=C\gamma_0^T\psi^*$. 
The charge  conjugation matrix $C$ has to satisfy $C\gamma_\mu C^{-1}=
-\gamma_\mu^T$, so that for instance the Dirac equation for a 
charged fermion in the presence of an electromagnetic field, $(i\dslash 
-e\Aslash-m)\psi=0$ implies that  $(i\dslash 
+e\Aslash-m)\psi^c=0$, i.e. that the antiparticle field has the same mass but 
 charges opposite to those of the particle field.
Since for a chiral projection one can show that $(\psi_L)^c=(P_L\psi)^c
=P_R\psi^c= (\psi^c)_R$, i.e. this conjugation changes the chirality of
the field, one has that $\psi^c$ is related to the $CP$ conjugate
of $\psi$. Notice  that $(\psi_L)^c$ describes exactly the same 
two degrees of freedom described by $\psi_L$, but somehow using a 
$CP$ reflected formalism. For instance for the neutrinos, the $\nu_L$ 
operator annihilates the left handed neutrino and creates the right
handed antineutrino, while the $(\nu_L)^c$ 
operator annihilates the right handed antineutrino and creates the left
handed neutrino.

We can then now write the advertised Majorana mass term, as
$$-\mathcal{L}_M={1\over 2}m[\overline{\nu_L}
(\nu_L)^c+\overline{(\nu_L)^c}\nu_L].$$
This mass term has the required Lorentz structure (i.e. the 
$L\leftrightarrow R$ transition) but one can see that it does not
preserve any $U(1)$ phase symmetry, i.e. it violates the so 
called lepton number by two units. 
If we introduce the Majorana field $\nu\equiv \nu_L+
(\nu_L)^c$, which under conjugation transforms into itself 
($\nu^c=\nu$), the mass term becomes just 
$\mathcal{L}_M=-m\bar\nu\nu/2$.

Up to now we have introduced the Majorana mass by hand, 
contrary to the case of the charged fermions where it arose from a
Yukawa coupling in a spontaneously broken theory. To follow the same
procedure with the neutrinos presents however 
a difficulty,  
because the standard model neutrinos belong to $SU(2)$ doublets,
and hence to write an electroweak singlet Yukawa coupling 
it is necessary to introduce an $SU(2)$ triplet  Higgs field
$\vec\Delta$ (something which is not particularly attractive). 
The coupling  $\mathcal{L}\propto \overline{L^c} \vec\sigma L\cdot 
\vec\Delta$ would then lead to the Majorana mass term after the 
neutral component of the scalar gets a VEV. 
 Alternatively,
the Majorana mass term could be a loop effect in models where the
neutrinos have lepton number violating couplings to new scalars, 
as in the so-called Zee models or in the supersymmetric models
with $R$ parity violation.
These models have as interesting features that the masses are 
naturally suppressed by the loop, 
and they are attractive also if one looks 
for scenarios where the neutrinos have relatively large dipole 
moments, since a photon can be attached to the charged 
particles in the  loop.

However, by far the nicest possibility to give neutrinos a mass is the
so-called see-saw model\cite{sees}. 
In this scenario, which naturally occurs in grand unified
models such as $SO(10)$, one introduces the $SU(2)$ singlet right
handed neutrinos. One has now not only the ordinary Dirac mass term,
but also a Majorana
mass for the singlets which is generated by the VEV of an $SU(2)$ singlet
Higgs, whose natural scale is the scale of breaking of the grand
unified group, i.e. in the range $10^{12}$--$10^{16}$~GeV. Hence the 
Lagrangian will contain
$$\mathcal{L}_M={1\over 2}\overline{(\nu_L,(N_R)^c)}\begin{pmatrix}
0&m_D\cr m_D&M\end{pmatrix}\begin{pmatrix}(\nu_L)^c\cr N_R\end{pmatrix}+
 h.c..
\label{seesaweq}$$
The mass eigenstates are two Majorana fields with masses $m_{light}
\simeq m_D^2/M$ and $m_{heavy}\simeq M$. Since $m_D/M\ll 1$, we 
see that $m_{light}\ll
m_D$, and hence the lightness of the known neutrinos  is here 
related to the 
heaviness of the sterile states $N_R$.

If we actually introduce one singlet neutrino per family, the
mass terms in eq.~(\ref{seesaweq}) are $3\times 3$ matrices. Notice 
that if $m_D$ is similar to the up-type quark masses, as happens in 
$SO(10)$, one would have $m_{\nu_3}\sim m_t^2/M\simeq 
0.04$~eV$(10^{15}$~GeV/$M$).  It is clear then that in these scenarios
the observation of neutrino masses near 0.1~eV would point out to 
new physics at about the GUT scale, while for $m_D\sim {\rm GeV}$ this would
correspond to singlet neutrino masses at an intermediate scale $M\sim
10^{10}$--$10^{12}$~GeV, also of great theoretical interest.

\subsection{Neutrino mixing and oscillations:}

If neutrinos are  massive, there is no reason to expect that the mass
eigenstates ($\nu_k$, with $k=1,2,3$) 
would coincide with the flavour (gauge) eigenstates ($\nu_\alpha$,
with $\alpha=e, \mu, \tau$, where we are adopting the flavor basis such that
for the charged leptons the flavor eigenstates coincide with the mass
eigenstates),  and hence, in the same way that quark states are mixed through
the Cabibbo, Kobayashi and Maskawa matrix, neutrinos would be related
through the Maki, Nakagawa and Sakita
\cite{ma62} (and Pontecorvo) mixing matrix, 
i.e. $\nu_\alpha=V^*_{\alpha k}\nu_k$. This matrix  can be parametrized as 
($c_{12}\equiv \cos\theta_{12}$, etc.) 
\begin{eqnarray*}
V=\begin{pmatrix}c_{12}c_{13} & c_{13}s_{12} & s_{13}e^{-i\delta}\cr
-c_{23}s_{12}-c_{12}s_{13}s_{23}e^{i\delta} & c_{12}c_{23}
- s_{12}s_{13}s_{23}e^{i\delta} & c_{13}s_{23}\cr
s_{23}s_{12}-c_{12}c_{23}s_{13}e^{i\delta} & -c_{12}
s_{23} -c_{23}s_{12}s_{13}e^{i\delta} & c_{13}c_{23}\end{pmatrix}
\begin{pmatrix}e^{i\frac{\alpha_1}{2}} & 0 & 0 \cr
0 & e^{i\frac{\alpha_2}{2}} & 0 \cr
0 & 0 & 1\end{pmatrix}.
\label{mns}
\end{eqnarray*}
The phases $\alpha_{1,2}$  here cannot be removed by a rephasing of the
fields (as is done for quarks) if the neutrinos are Majorana particles, since
such rotations would then introduce a complex phase in the neutrino masses.

The possibility that neutrino flavour eigenstates be a superposition
of mass eigenstates allows for the
phenomenon of neutrino oscillations.  This is a quantum mechanical
interference effect (and as such it is sensitive to quite small masses)
and arises because different mass eigenstates propagate differently,
and hence the flavor composition of a state can change with time.

To see this consider a flavour eigenstate neutrino $\nu_\alpha$ with
momentum $p$ produced at time $t=0$ (e.g. a $\nu_\mu$ produced in the
decay $\pi^+\to \mu^++\nu_\mu$). The initial state is then
$$|\nu_\alpha\rangle =\sum_kV^*_{\alpha k}|\nu_k\rangle .$$
We know that the mass eigenstates evolve with time according to 
 $|\nu_k(t,x) \rangle={\rm exp}[i(px-E_kt)]|\nu_k\rangle$. In the
relativistic limit relevant for neutrinos, one has that
$E_k=\sqrt{p^2+m_k^2}\simeq p+m^2_k/2E$, and thus the different mass
eigenstates will acquire different phases as they propagate.  Hence,
the probability of observing a flavour $\nu_\beta$ at time $t$ is just
$$ P(\nu_\alpha\to\nu_\beta)=|\langle \nu_\beta|\nu(t)\rangle|^2=
|\sum_kV^*_{\alpha k}e^{-i{m_i^2\over 2E}t}V_{\beta k}|^2.$$
Taking into account the explicit expression for $V$, 
it is easy to convince oneself that the
Majorana phases $\alpha_{1,2}$  do not enter into the oscillation
probability, and hence oscillation phenomena cannot tell whether neutrinos are
Dirac or Majorana particles. 

In the case of two generations,  $V$ can be taken just as a rotation
$R_\theta$ with
mixing angle $\theta$, so that one has
$$ P(\nu_\alpha\to\nu_\beta)=\sin^22\theta\ \sin^2\left( {\Delta
m^2x\over 4E}\right),$$
which depends on the squared mass difference $\Delta
m^2=m_2^2-m_1^2$, since this is what gives the phase difference in the
propagation of the mass eigenstates. Hence, the amplitude of the
flavour oscillations is given by $\sin ^22\theta$ and the oscillation
length of the modulation is $L_{osc}\equiv 4\pi E/ \Delta
m^2\simeq 2.5$~m $E$[MeV]/$\Delta m^2$[eV$^2$]. We see then that
neutrinos will typically oscillate with a macroscopic wavelength. For
instance, putting a detector at $\sim 1$~km from a reactor (such as in the
CHOOZ experiment) allows to
test oscillations of $\nu_e$'s to another flavour (or into a singlet
neutrino) down to $\Delta m^2\sim 10^{-3}$~eV$^2$ if sin$^22\theta$
is not too small ($\geq 0.1$). 
These kind of experiments look essentially for the disappearance of
the reactor $\nu_e$'s, i.e. to a reduction in the original $\nu_e$
flux. When one uses neutrino beams from accelerators, it
becomes possible also to study the disappearance of muon neutrinos into
another flavor, and also the appearance of a flavour different
from the original one, with the advantage that one becomes sensitive
to very small oscillation amplitudes (i.e. small sin$^22\theta$
values), since the observation of only a few events is enough to
establish a positive signal. At present there is one experiment (LSND)
claiming a positive signal of $\nu_\mu\to\nu_e$ conversion, but this highly
suspected result is expected to be clarified unambiguously by the MINIBOONE
experiment at Fermilab during 2005. 
 The appearance of $\nu_\tau$'s out of a $\nu_\mu$
beam was searched by CHORUS and NOMAD at CERN without success, but these
experiments were only sensitive to squared mass differences larger than $\sim
{\rm eV}^2$.
There are two experiments which have obtained recently solid
evidence of neutrino oscillations (K2K and Kamland), but let's however,
following the historical evolution, start with the discussion of solar and
atmospheric neutrinos and the clues they have given in favor of non-vanishing
neutrino masses.

\subsection{Solar neutrinos and oscillations in matter:}

The Sun gets its energy from the fusion reactions taking place in its
interior, where essentially four protons combine to 
form a He nucleus. By charge
conservation this has to be accompanied by the emission of two
positrons and, by lepton number conservation in the weak processes, two
$\nu_e$'s have to be produced. This fusion liberates 27~MeV of
energy, which is eventually emitted mainly (97\%) as photons and the
rest (3\%) as neutrinos. Knowing the energy flux of the solar
radiation reaching us ($k_\odot\simeq 1.5$~kW/m$^2$), it is
then simple to estimate that the solar neutrino flux at Earth is
$\Phi_\nu\simeq 2k_\odot/27$~MeV $\simeq 6\times
10^{10}\nu_e/$cm$^2$s, which is a very large number indeed. Since there are
many possible paths for the four protons to lead to an He nucleus, the solar
neutrino spectrum consists of different components: the so-called 
$pp$ neutrinos are the
more abundant, but have very small energies ($<0.4$~MeV), the $^8B$ neutrinos 
are the
more energetic ones ($<14$~MeV) but are much less in number, there are also
some  monochromatic lines ($^7Be$ and $pep$ neutrinos), and then the $CNO$ 
and $hep$ neutrinos.

Many experiments  have looked for these solar neutrinos: the
radiochemical experiments with $^{37}$Cl at Homestake and with gallium
at SAGE, GALLEX and GNO, and the water Cherenkov real time detectors
(Super-) Kamiokande and more recently the heavy water Subdury Neutrino
Observatory (SNO)\footnote{See the Neutrino 2004 homepage
at http://neutrino2004.in2p3.fr for this year's results.}. The
result which has puzzled physicists for almost thirty years is
that only between 1/2 to 1/3 of the expected fluxes were
observed. Remarkably, Pontecorvo \cite{po67} noticed even before the
first observation of solar neutrinos by Davies that neutrino
oscillations could reduce the expected rates. We note that the
oscillation length of solar neutrinos ($E\sim 0.1$--10~MeV) is of the
order of 1 AU for $\Delta m^2\sim 10^{-11}$~eV$^2$, and hence even
those tiny neutrino masses could have had observable effects if the mixing
angles were large (this would be the `just so' solution to the solar
neutrino problem). Much more interesting became the possibility of
explaining the puzzle by resonantly enhanced oscillations of neutrinos
as they propagate outwards through the Sun. Indeed, the solar medium
affects $\nu_e$'s differently than $\nu_{\mu,\tau}$'s (since only the
first interact through charged currents with the electrons present),
and this modifies the oscillations in a beautiful way through an
interplay of neutrino mixings and matter effects, in the so called MSW
effect \cite{msw}. 

To see how this effect works it is convenient to write the effective CC
interaction (after a Fierz rearangement) as
\begin{equation}
H^{CC}=\sqrt{2}G_F\bar e\gamma_\mu(1-\gamma_5)e{\bar \nu}_{eL}\gamma^\mu
\nu_{eL}.
\end{equation}
Since the electrons in a normal medium (such as the Sun or the interior of the
Earth) are non-relativistic, one can see that
$\bar e \gamma_\mu e\to (N_e,\vec 0)$, where $N_e$ is the electron density,
while for the axial vector part one gets
$\bar e\gamma_\mu\gamma_5 e\to(0,\vec S_e)$, which vanishes for an unpolarised
medium, as is the case of interest here. 
 This means that the electron neutrinos will feel a potential 
\begin{equation}
V_{CC}=\langle e\nu_e|H^{CC}|e\nu_e\rangle \simeq \sqrt{2}G_FN_e
\end{equation}
(and for the antineutrinos the potential will have a minus sign in front).
The evolution of the neutrino states will hence be determined by a
Schroedinger like equation of the form (for the case of just two flavor
mixing, i.e. $\alpha=\mu$ or $\tau$)
\begin{equation}
i\frac{\rm d}{{\rm d}t}\begin{pmatrix}\nu_e\cr\nu_\alpha\end{pmatrix}=
\left\{R_\theta\left[\begin{matrix}p+\frac{m_1^2}{2E} & 0\cr 0& p+\frac{m_2^2}
{2E}\end{matrix}
\right] R_\theta^T+\left[\begin{matrix}\sqrt{2}G_FN_e&0\cr
      0&0\end{matrix}\right]
+NC\right\}\begin{pmatrix}\nu_e\cr \nu_\alpha\end{pmatrix}
\end{equation}
where $\theta$ is the vacuum mixing angle and
\begin{equation}
R_\theta\equiv \begin{pmatrix} \cos\theta&\sin\theta\cr
-\sin\theta&\cos\theta\end{pmatrix}.
\end{equation}
The terms indicated as $NC$ correspond to the effective potential induced by
the neutral current interactions of the neutrinos with the medium, but since
these are flavor blind, this term is proportional to the identity matrix and
hence does not affect the flavor oscillations, so that we can ignore it in the
following (these terms could be relevant e.g. when studying oscillations into
sterile neutrinos, which unlike the active ones do not feel NC
interactions). 

To solve this equation it is convinient to introduce a mixing angle in matter
$\theta_m$ and define the neutrino matter eigenstates
\begin{equation}
\begin{pmatrix}\nu_1^m\cr\nu_2^m\end{pmatrix}
\equiv R^T_{\theta_m}\begin{pmatrix}\nu_e\cr\nu_\alpha\end{pmatrix}
\end{equation}
such that the evolution equation becomes
\begin{equation}
i\frac{\rm d}{{\rm d}t}\begin{pmatrix}\nu_e\cr\nu_\alpha\end{pmatrix}=
\left\{\frac{1}{4E}R_{\theta_m}\left[\begin{matrix}-\Delta\mu^2 & 0\cr 0& 
\Delta\mu^2\end{matrix}
\right] R_{\theta_m}^T+\lambda I\right\}
\begin{pmatrix}\nu_e\cr \nu_\alpha\end{pmatrix}
\end{equation}
where the diagonal term $\lambda I$ is again irrelevant for oscillations.
It is simple to show that to have such `diagonalisation' of the effective
Hamiltonian one needs
\begin{equation}
\Delta\mu^2=\Delta m^2\sqrt{(a-c2\theta)^2+s^22\theta}
\end{equation}
\begin{equation}
s^2{\theta_m}={s^22\theta\over (c2\theta-a)^2+s^22\theta}
\end{equation}
where $a\equiv 2\sqrt{2}G_FN_eE_\nu/\Delta m^2$ is just the ratio between the
effective CC matter potential and the energy splitting between the two vacuum
mass eigenstates. It is clear then that there will be a resonant behaviour for
$a=c2\theta$, i.e. for
\begin{equation}
\Delta m^2 c2\theta=2\sqrt{2}G_FN_e
E_\nu\simeq \left(\frac{Y_e}{0.5}\right) \left(\frac{E_\nu}{10\ {\rm
    MeV}}
\right) {\rho \over 100\ {\rm g/cm}^3}10^{-4}\ {\rm eV}^2,
\end{equation}
where for the second relation we used that $N_e= Y_e\rho/m_p$, with
$Y_e\equiv N_e/(N_n+N_p)$ (for the Sun $Y_e\sim 0.7$--0.8).
One can see that at the resonance the matter mixing angle becomes maximal,
i.e. $\theta_m=\pi/4$, while at densities much larger than the resonant one it
becomes $\sim \pi/2$ (i.e.  one gets 
for densities much larger than the resonance one
that $\nu_e\simeq \nu_2^m$).

In the case of the Sun, one has that the density decreases approximately
exponentially
\begin{equation}
\rho\simeq 10^2 \frac{\rm g}{\rm cm^3}\exp(-r/h)
\end{equation}
with the scale height being $h\simeq 0.1R_\odot$ in terms of the solar
radius  $R_\odot$. Hence, solar neutrinos, which are produced near the center, 
will cross a resonace in their way out only if $\Delta m^2<10^{-4}$~eV$^2$,
and moreover only if $\Delta m^2>0$. For the opposite sign of $\Delta m^2$ only
antineutrinos, which are however not produced in fusion processes, could meet
a resonance in normal matter\footnote{Alternatively, one can stick to positive
  values of $\Delta m^2$ and consider vacuum mixing angles in the range $0\leq
  \theta\leq \pi/2$, with the range $\pi/4\leq \theta\leq \pi/2$ sometimes
  called the `dark side' of the parameter space.}.

One can also associate a width $\delta_R$ to the resonance, 
corresponding to the
density for which $|a-c2\theta|\simeq s2\theta$, i.e. $|da/dr|_R\delta_R\simeq
s2\theta$. This leads to $\delta_R\simeq h\ {\rm tg} 2\theta$.
This width is useful to characterize the two basic regimes of resonant flavor
conversions, which are the adiabatic one, taking place when the oscillation
length in matter is much smaller than the resonance width,  i.e. for
\begin{equation}
{4\pi E_\nu\over \Delta \mu^2|_R}={4\pi E_\nu\over \Delta 
m^2s2\theta}<h\ {\rm
  tg}2\theta,
\end{equation}
and the opposite one, which is called non-adiabatic, for which the
resonance is so narrow that the oscillating neutrinos effectively don't see it
and hence no special flavor transition occurs at the resonant crossing.
The adiabatic condition can be rewritten as
\begin{equation}
{s^22\theta\over c2\theta}>\left(\frac{E_\nu}{10\ {\rm
    MeV}}\right)\left(\frac{\rm 6\times 10^{-8}eV^2}{\Delta m^2}\right).
\end{equation}

To better understand the flavor transition during resonance crossing, it proves
convenient to write down the evolution equation for the matter mass
eigenstates, which is easily obtained as (ignoring terms proportional to the
identity) 
\begin{equation}
i\frac{\rm d}{{\rm d}x}\begin{pmatrix}\nu_1^m\cr\nu_2^m\end{pmatrix}
=\begin{pmatrix}-\frac{\Delta \mu^2}{4E} & -i\frac{{\rm d}\theta_m}{{\rm
    d}x}\cr
i\frac{{\rm d}\theta_m}{{\rm d}x}&\frac{\Delta \mu^2}{4E}\end{pmatrix}
\begin{pmatrix}\nu_1^m\cr\nu_2^m\end{pmatrix}
\label{mswmatter}
\end{equation}
We see that in the adiabatic case, the off diagonal terms in this equation 
are negligible
and hence during the resonance crossing the matter mass eigenstates remain
themselves so that the flavor of the neutrinos changes just  following the
change on the matter mixing angle with the varying electron density. This
adiabatic behaviour is also relevant for the propagation of neutrinos in a
medium of constant density (as is sometimes a good approximation for the
propagation  through the Earth), and in this case the matter effects just
change the mixing angle and the frequency of the oscillations among neutrinos.
When the propagation is non-adiabatic, the off-diagonal terms 
in Eq.~(\ref{mswmatter}) induce transitions between the different matter mass
eigenstates as the resonance is crossed. Indeed the probability of jumping
from one eigenstate to the other during resonance crossing for an exponential
density profile can be written as
\begin{equation}
P_c(\nu_1^m\to\nu_2^m) = {\exp{(-\gamma\sin^2\theta)}-\exp{(-\gamma)}\over 1-\exp{(-\gamma)}},
\end{equation}
where the adiabaticity parameter is 
$\gamma\equiv \pi h\Delta m^2/E$ (notice that
for an electron density varying in a more general way, not just exponentially,
it is usually a good approximation to  replace $h$ in the above formulas
by $|(dN_e/dr)/N_e|^{-1}_R$).

The many observations of the solar 
 neutrino fluxes by the different experiments, which having
different energy thresholds are sensitive to the oscillation probabilities in
different energy ranges (moreover water Cherenkov detectors can measure the
neutrino spectrum directly above their thresholds), 
and also the non-observation of a possible diurnal modulation induced 
by the matter
effects when neutrinos have to cross the Earth before reaching the detectors
during night-time observations, have converged over the years towards the
so-called large mixing angle (LMA) solution as the one required to account for
the solar neutrino observations. This one corresponds to mixing of $\nu_e$
with some combination of $\nu_\mu$ and $\nu_\tau$ flavors involving a mass
splitting between the mass eigenstates of\footnote{this number
  includes also the results from the Kamland experiment\cite{kamland}.}
 $\Delta m^2_{sol}=+(7.9^{+0.6}_{-0.5})\times 10^{-5}$~eV$^2$ 
with a mixing angle given by $\tan^2\theta_{sol} =
0.40^{+0.10}_{-0.07}$.
 This values imply that the resonance layer is actually at large
densities near the center of the Sun, and that it is quite wide, so that
matter oscillations are well in the adiabatic regime.

Another crucial result obtained in 2002 was the independent measurement
of the CC and NC interactions of the solar neutrinos with the 
heavy water experiment SNO\cite{sno}. 
The result is that the NC rates, which are
sensitive to the three flavors of active neutrinos, are indeed consistent with
the solar model expectations for $\nu_e$ alone in the absence of oscillations,
while the CC rates, which are sensitive to the electron neutrinos alone, show
the deficit by a factor $\sim 3$, indicating that the oscillations have
occured and that they convert electron neutrinos  into other 
active neutrino flavors ($\nu_{\mu,\tau}$).

\begin{figure}[t]
\centerline{\hbox{\epsfxsize=7cm \epsfbox{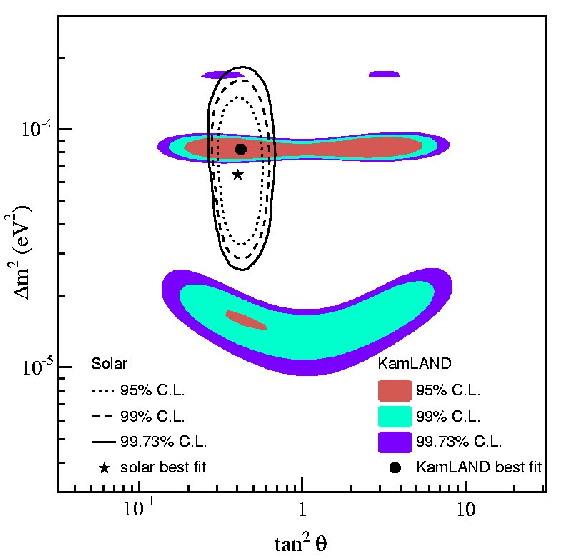} }}
\caption{\footnotesize  Bounds from the KAMLAND experiment. The region 
favored by solar neutrino observations is that with unfilled contours.
\label{kamland}}
\end{figure}

The last remarkable result that has confirmed this picture has been the
observation of oscillations of reactor neutrinos (from a large number of
japanese reactors) using a huge 1 kton scintillator detector (KAMLAND),
measuring oscillations over distances of $\sim 10^2$~km, and the reduction
found from expectations just agrees with those resulting from the LMA
parameters, and have actually restricted the mass splitting involved to the
narrow range just mentioned, as is shown in figure~\ref{kamland} (from
the Kamland experiment).

\subsection{Atmospheric neutrinos:}

\begin{figure}[t]
\centerline{\hbox{\epsfxsize=7cm \epsfbox{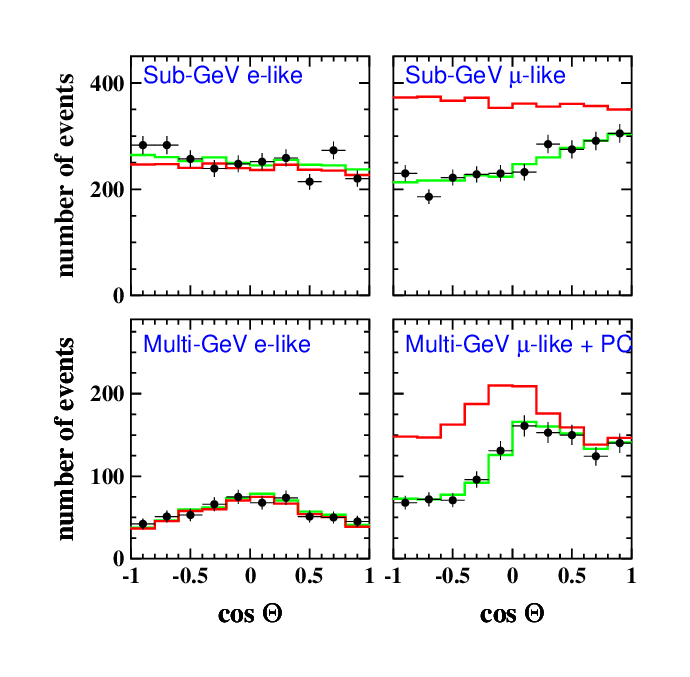} }}
\caption{\footnotesize  Distribution of the contained and partially
contained event data versus cosine of the zenith angle
($\cos\theta=-1$ being up-going, while $+1$ being down-going) for two
energy ranges,  from  Super-Kamiokande data. The solid line corresponds to
the expectations with no oscillations, while the lighter line is for
$\nu_\mu\to\nu_\tau$ oscillations with maximal mixing and $\Delta
m^2=0.003$~eV$^2$.
\label{ud}}
\end{figure}

When a cosmic ray  (proton or nucleus) hits the atmosphere and knocks a
nucleus a few tens of km above ground, an hadronic (and
electromagnetic) shower is initiated, in which pions in particular are
copiously produced. The charged pion decays are the main source of
atmospheric neutrinos through the chain $\pi\to \mu\nu_\mu\to
e\nu_e\nu_\mu\nu_\mu$. One expects then twice as many $\nu_\mu$'s than
$\nu_e$'s  (actually at very high energies, $E_\nu\gg$ GeV, the parent
muons may reach the ground and hence be stopped before decaying, so
that the expected ratio $R\equiv
(\nu_\mu+\bar\nu_\mu)/(\nu_e+\bar\nu_e)$ should be even larger than
two at high energies). However, the observation of the atmospheric
neutrinos by e.g. IMB, Kamioka, Soudan, MACRO and Super-Kamiokande
indicates that there is a deficit of muon neutrinos, with
$R_{obs}/R_{th}\simeq 2/3$. 

More remarkably, 
the Super-Kamiokande experiment\cite{superk} observes a zenith angle dependence
indicating that neutrinos coming from above (with pathlengths $d\sim
20$~km) had not  enough time to oscillate, especially in the multi-GeV
sample for which the neutrino
 oscillation length is larger,  while those from below
($d\sim 13000$~km) have already oscillated (see figure~\ref{ud}). 
The most plausible explanation for
these effects is an oscillation $\nu_\mu\to\nu_\tau$ with maximal
mixing 
$\sin^22\theta_{atm}= 1.00\pm 0.04$ and $\Delta m^2\simeq (2.5\pm 0.4)\times
10^{-3}$~eV$^2$,
and as shown in
fig.~\ref{ud} (from the Super-Kamiokande experiment) 
the fit to the observed angular dependence
is in excellent agreement with the oscillation hypothesis. Since the
electron flux shape is in good agreement with the theoretical
predictions\footnote{The theoretical uncertainties in the absolute
flux normalisation may amount to $\sim 25$\%, but the predictions for
the ratio of muon to electron neutrino flavours and for their angular
dependence are much more robust.}, this means that the oscillations
from $\nu_\mu\to\nu_e$ can not provide a satisfactory explanation for
the anomaly (and furthermore they are also excluded from the negative
results of the CHOOZ reactor search for oscillations).  On the other
hand, oscillations to sterile states would be affected by matter
effects ($\nu_\mu$ and $\nu_\tau$ are equally affected by neutral
current interactions when crossing the Earth, while sterile states are
not), and this would modify the angular dependence of the oscillations
in a way which is not favored by observations.
The oscillations into active states ($\nu_\tau$) is also favored by
observables which depend on the neutral current interactions, such as
the $\pi_0$ production in the detector or the `multi ring' 
events.

An important experiment which  has confirmed  the
oscillation solution to the atmospheric neutrino anomaly is K2K\cite{k2k}, 
consisting of a beam of muon neutrinos sent from KEK to the
Super-Kamiokande detector (baseline of 250~km). The 
results indicate that there is a deficit of muon
neutrinos at the detector ($150.9\pm 10$ events 
expected with only 108
observed), consistent with the expectations from the oscillation
solution. 

 It is remarkable that the mixing angle involved seems to be
maximal, and this, together with the large mixing angle required in 
the solar sector,
is giving fundamental information about the new physics underlying the origin
of the neutrino masses, which seem to be quite different from what is observed
in the quark sector.

\subsection{The direct searches for the neutrino mass:}

Already in his original paper on the theory of weak interactions Fermi
had noticed that the observed shape of the electron spectrum was
suggesting a small mass for the neutrino. The sensitivity to
$m_{\nu_e}$ in the decay $X\to X'+e+\bar\nu_e$ arises clearly because the
larger $m_\nu$, the less available kinetic energy remains for the
decay products, and hence the maximum electron energy is reduced. To
see this consider the phase space factor of the decay, $d\Gamma\propto
d^3 p_ed^3p_\nu\propto p_eE_edE_ep_\nu E_\nu dE_\nu \delta(E_e+E_\nu
-E_0)$, with $E_0$ being the total energy available for the leptons 
in the decay: $E_0\simeq M_X-M_{X'}$ (neglecting the nuclear recoil). 
This leads to a differential electron
spectrum proportional to $d\Gamma /dE_e\propto p_eE_e(E_0-E_e)
\sqrt{(E_0-E_e)^2-m_\nu^2}$, whose shape near the endpoint ($E_e\simeq
E_0-m_\nu$) 
depends on $m_\nu$ (actually the slope becomes infinite at the
endpoint for $m_\nu\neq 0$, while it vanishes for $m_\nu=0$).

Since the fraction of events in an interval $\Delta E_e$ around the
endpoint is $\sim (\Delta E_e/Q)^3$ (where $Q\equiv E_0-m_e$), 
to enhance the sensitivity to the
neutrino mass it is better to use processes with small $Q$-values,
what makes the tritium the most sensitive nucleus
($Q=18.6$~keV). Experiments at Mainz and Troitsk have allowed
to set the bound $m_{\nu_e}\leq 2.2$~eV\cite{mainz}.
It is important to keep in mind that in the presence of flavor mixing, as is
indicated by solar and atmospheric neutrino observations, the bound from beta
decays actually applies to the quantity $m_\beta\equiv\sqrt{\sum |V_{ei}|^2
  m_i^2}$, since the beta spectrum will actually be an incoherent
supperposition of spectra (weighted by $|V_{ei}|^2$) with different endpoints,
but which are however not resolved by the experimental apparatus. Hence, given
the constraints we already have on the mixing angles, and the mass splittings
observed, these results already constrain significantly all three neutrino
masses.

Regarding the muon neutrino, a direct bound on its mass can be set by
looking to its effects on the available energy for the muon in the
decay of a pion at rest, $\pi^+\to\mu^++\nu_\mu$. From the knowledge
of the $\pi$ and $\mu$ masses, and measuring the momentum of the
monochromatic muon, one can get the neutrino mass through the relation 
$$m^2_{\nu_\mu}=m^2_\pi +m^2_\mu-2m_\pi\sqrt{p^2_\mu+m^2_\mu}.$$
The best bounds at present are $m_{\nu_\mu}\leq 170$~keV from PSI, and
again they are difficult to improve through this process since the
neutrino mass comes from the difference of two large quantities. There
is however a proposal to use  the muon $(g-2)$ experiment at BNL to
become sensitive down to $m_{\nu_\mu}\leq 8$~keV.

Finally, the direct bound on the $\nu_\tau$ mass is $m_{\nu_\tau}\leq 17$~MeV
and comes from the effects it has on the available phase space of the
pions in the decay $\tau\to 5\pi+\nu_\tau$ measured at LEP.

\begin{figure}[t]
\centerline{\hbox{\epsfxsize=10cm \epsfbox{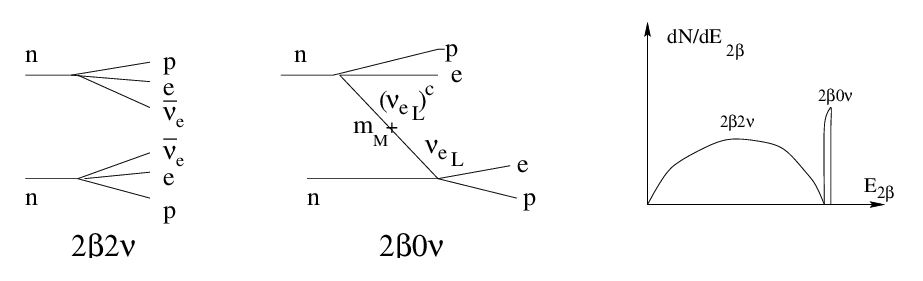} }}
\caption{\footnotesize 
Double beta decay with and without neutrino emission, and
qualitative shape of the expected added spectrum of the two electrons. 
\label{2beta}}
\end{figure}

To look for the electron neutrino mass, besides the endpoint of the
ordinary beta decay there is another interesting process, but which is
however only sensitive to a Majorana (lepton number violating)
mass. This is the so called double beta decay. 
Some nuclei can undergo transitions in which two beta decays take
place simultaneously, with the emission of two electrons and two
antineutrinos ($2\beta 2\nu$ in fig.~\ref{2beta}). These transitions
have been observed in a few isotopes ($^{82}$Se, $^{76}$Ge,
$^{100}$Mo, $^{116}$Cd, $^{150}$Nd) in which the single beta decay is
forbidden, and the associated lifetimes are huge
($10^{19}$--$10^{24}$~yr). However, if neutrinos were Majorana
particles, the virtual antineutrino emitted in one vertex could flip
chirality by a mass insertion and be absorbed in the second vertex as
a neutrino, as exemplified in fig.~\ref{2beta} ($2\beta 0\nu$). In
this way only two electrons would be emitted and this could be
observed as a monochromatic line in the added spectrum of the two
electrons. The non observation of this effect has allowed to set the
bound $m_{\nu_e}^{Maj}\equiv |\sum V_{ei}^2 m_i| \leq $~eV (by the Heidelberg--Moscow
collaboration at Gran Sasso). A reanalysis of the results of this experiment
even suggest a mass in the range 0.2--0.6~eV, but this controversial claim  is
expected to be reexplored by the next generation of double beta decay
experiments (such as CUORE).
There are even projects to improve the
sensitivity of $2\beta 0\nu$ down to $m_{\nu_e}\sim 10^{-2}$~eV, and
we note that this is quite relevant since as we have seen, if
neutrinos are indeed massive, it is somehow theoretically favored
(e.g. in the see saw models) that they are Majorana particles.

%

\end{document}